\documentstyle[epsfig]{aipproc}

\newcommand{\nub}{\overline{\nu}}
\newcommand{\rms}{\rm\scriptstyle}
\newcommand{\rmt}{\rm\textstyle}
\begin{document}
\title{Electroweak Physics at NuTeV}

\author{G.P. Zeller\\ for the NuTeV Collaboration}
\address{Northwestern University, Evanston IL 60208}

\maketitle

\begin{abstract}
The NuTeV experiment at Fermilab presents a determination of the electroweak 
mixing angle. High purity, large statistics samples of $\nu_\mu N$ and 
$\nub_\mu N$ events allow the use of the Paschos-Wolfenstein relation, 
a technique which considerably reduces systematic errors associated with 
charm production and other sources. Within the Standard Model, this 
measurement of $\sin^{2}\theta_{W}$ indirectly determines the W boson 
mass to a precision comparable to direct measurements from high energy 
$e^+e^-$ and $p\bar{p}$ colliders. NuTeV measures 
${\rmt sin^{2}\theta_{W}}^{({\rms on-shell)}}=
0.2253\pm0.0019({\rmt stat.})\pm0.0010({\rmt syst.})$, which implies 
$M_W=80.26\pm0.11~$ GeV. Outside the Standard Model, this result
can be used to explore the possibility of new physics; in particular, 
we present limits on both neutrino oscillations and the presence of extra 
neutral vector gauge bosons.
\end{abstract}

In deep inelastic neutrino-nucleon scattering, the weak mixing angle can
be extracted from the ratio of neutral current (NC) to charged current (CC)
total cross sections. A method for determining $\sin^2\theta_W$ that 
is much less dependent on sources of model uncertainty and the the details 
of charm production (the largest source of uncertainty in the previous 
neutrino measurement \cite{ccfr}) employs the Paschos-Wolfenstein 
relation \cite{pw}:

\begin{equation}
R^{-} 
= \frac{\sigma^{\nu}_{NC}-\sigma^{\bar \nu}_{NC}}
       {\sigma^{\nu}_{CC}-\sigma^{\bar \nu}_{CC}}
= \frac{R^{\nu}-rR^{\nub}}{1-r}=\frac{1}{2}-\sin^2\theta_W
\end{equation}

\noindent
where $R^{\nu,\bar \nu}=
\sigma^{\nu,\bar \nu}_{NC}/\sigma^{\nu,\bar \nu}_{CC}$, and 
$r=\sigma^{\bar \nu}_{CC}/\sigma^{\nu}_{CC}$. Unfortunately, the substantially
reduced uncertainties come at a price: R$^{-}$ is a more difficult 
quantity to measure experimentally because neutral current neutrino and 
antineutrino events have identical observed final states. The two samples 
can only be separated by knowing the incoming neutrino beam type. \\

High-purity neutrino and antineutrino beams were provided by the Sign Selected 
Quadrupole Train (SSQT) at the Fermilab Tevatron during the 1996-1997
fixed target run. Neutrinos are produced from the decay of pions and kaons 
resulting from interactions of 800 GeV protons in a BeO target. Dipole magnets
immediately downstream of the proton target bend pions and kaons of specified 
charge in the direction of the NuTeV detector, while wrong-sign and neutral 
mesons are stopped in beam dumps. The resulting beam is almost purely 
neutrino or antineutrino depending on the selected sign of the parent mesons
(opposite particle contamination is $\sim$0.1\%). In addition, the beam is 
almost purely muon neutrinos with a small ($\sim$1\%) contamination of 
electron neutrinos.

Neutrino interactions are then observed in the NuTeV detector \cite{detector}, 
which is located approximately 1.5 km downstream of the proton target. 
The detector consists of an 18m long, 690 ton steel-scintillator target 
followed by an instrumented iron-toroid spectrometer. 
The target calorimeter is composed of 168 3m x 3m x 5.1cm steel plates 
interspersed with liquid scintillation counters and drift chambers. 
The scintillation counters provide triggering information as well as a 
determination of the longitudinal event vertex, event length and visible 
energy deposition. The mean position of hits in the drift chambers help 
establish the transverse event vertex. The toroidal spectrometer, which 
determines muon sign and momentum, is not directly used for this analysis.
In addition, because the detector was continuously calibrated through 
exposure to a wide energy range of test beam hadrons, electrons and muons, 
many systematics related to detector effects were substantially reduced.

\section*{Within the Standard Model}        

In order to measure $\sin^2\theta_W$, observed neutrino events must be 
separated into charged current (CC) and neutral current (NC) categories. 
Both CC and NC neutrino interactions initiate a cascade of hadrons in the 
target that is registered in both the scintillation counters and drift 
chambers. However, muon neutrino CC events are distinguished by the presence 
of a final state muon, which typically penetrates well beyond the hadronic 
shower and deposits energy in a large number of consecutive scintillation 
counters. These differing event topologies enable the statistical separation 
of CC and NC interactions based solely on event length (\emph{i.e.}, 
on the presence or absence of a muon in an event). Events with a long length 
(spanning more than 20 counters) are identified as CC candidates; those with a 
short length (spanning less than 20 counters) as NC candidates. The 
experimental quantity measured in both neutrino and antineutrino modes is 
the ratio:

\begin{equation}
{\rmt R_{meas}} = \frac{\# \hspace{0.05in} \rmt SHORT \hspace{0.05in} events}
                       {\# \hspace{0.05in} \rmt LONG \hspace{0.05in} events}
              = \frac{\# \hspace{0.05in} \rmt NC  \hspace{0.05in} candidates}
                     {\# \hspace{0.05in} \rmt CC \hspace{0.05in} candidates}
\end{equation}

The ratios of short to long events ({$\rmt R_{meas}$}) measured in the NuTeV 
data are 0.4198 $\pm$ 0.0008 in the neutrino beam and 0.4215 $\pm$ 0.0017 
in the antineutrino beam. A Standard Model value of $\sin^2\theta_W$ can be 
directly extracted from these measured ratios by using a detailed Monte Carlo 
simulation of the experiment. The Monte Carlo must include the integrated 
neutrino fluxes, the neutrino cross section, and a detailed description
of the NuTeV detector. More detailed information on the specific components 
of the Monte Carlo simulation can be found elsewhere \cite{dpfproc}.

Using our separate high-purity neutrino and antineutrino data sets, NuTeV 
measures the following linear combination of $R^{\nu}$ and $R^{\nub}$:

\begin{equation}
     R^{-} = R^{\nu}-x R^{\nub}
\end{equation}

\noindent
The value for {\em{x}} is selected using the Monte Carlo in order to minimize 
uncertainties related to charm quark production. The single remaining free 
parameter in the Monte Carlo, $\sin^2\theta_W$, is then varied until the 
model calculation of $R^{-}$ agrees with what is measured in the data. The 
preliminary result from the NuTeV data sample for $M_{\rms top}$=175 GeV and 
$M_{\rms Higgs}$=150 GeV is:

\begin{equation}
    \sin^2\theta_W^{({\rms on-shell)}}=0.2253\pm0.0019({\rmt stat.})\pm0.0010({\rmt syst.})
\end{equation}

\noindent
Having chosen the convention
$\sin^2\theta_W^{({\rms on-shell)}}\equiv 1- \frac{M_{W}^2}{M_{Z}^2}$, 
and given the well-determined Z mass from LEP, our result implies:

\begin{equation}
    M_{W}=80.26\pm0.10({\rmt stat.})\pm0.05(\rmt syst.)
         =80.26\pm0.11 GeV
\end{equation}

\begin{figure}[ht]	
\mbox{
\begin{minipage}{0.5\textwidth}
\centerline{\epsfxsize 2.0 truein \epsfbox{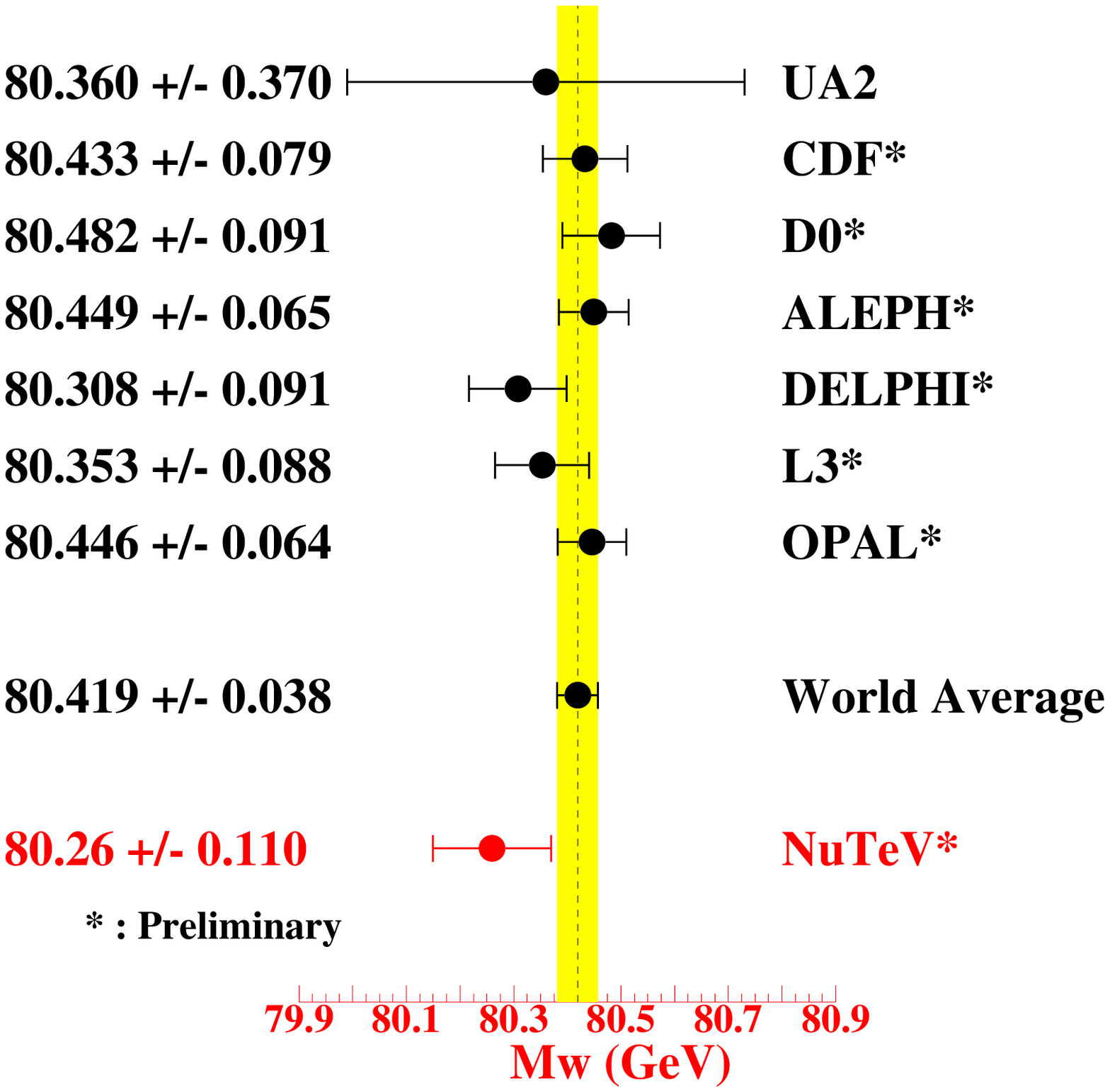}}   
\caption[]{Direct W boson mass measurements compared with this result.}
\label{fig:mw}
\end{minipage}\hspace*{0.02\textwidth}	
\begin{minipage}{0.5\textwidth}
\centerline{\epsfxsize 2.0 truein \epsfbox{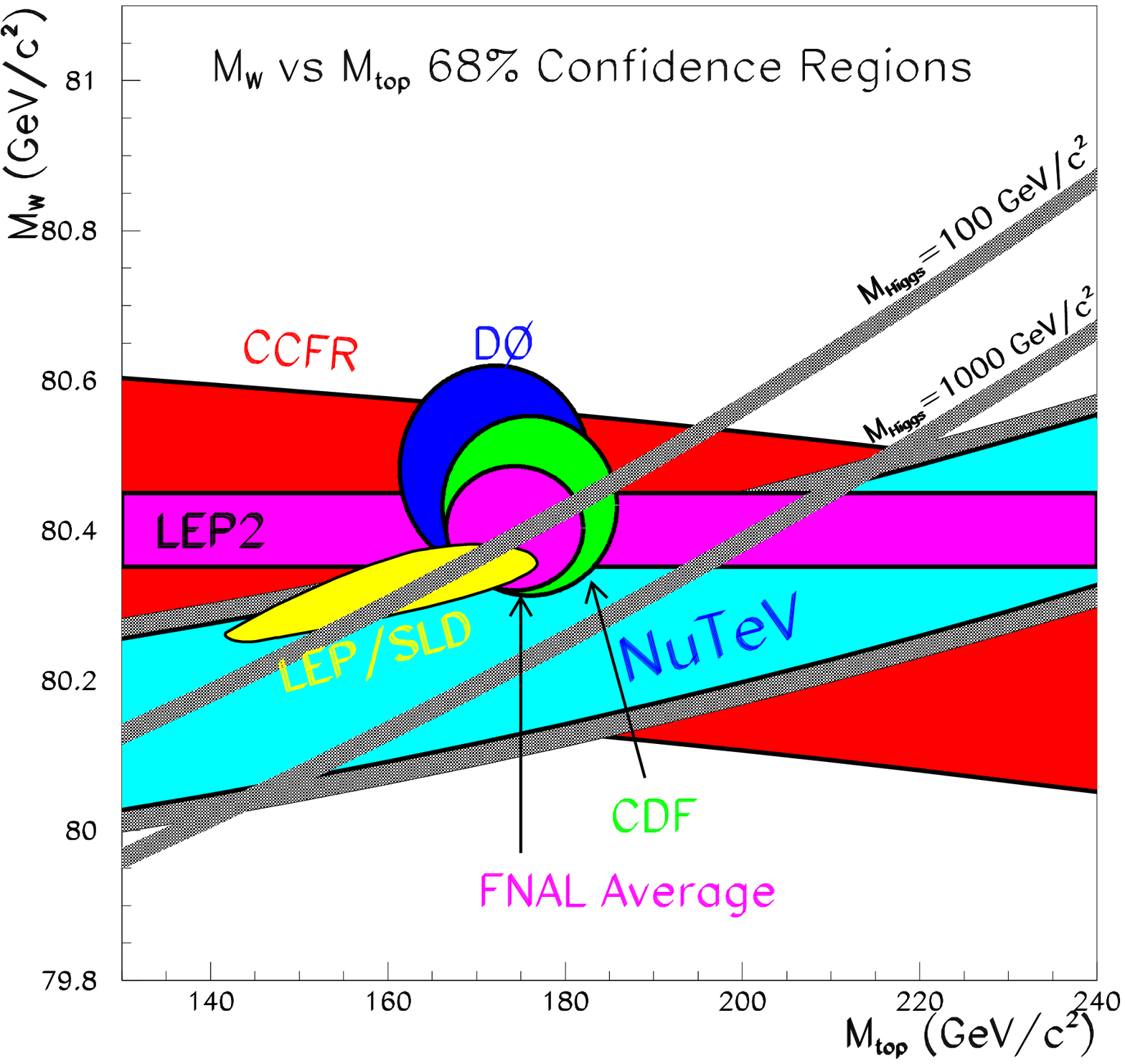}}   
\caption[]{Experimental constraints presented on the $M_{W}$-$M_{\rms top}$ 
           plane. The two narrow bands indicate the Standard Model predictions
           for $M_{\rms Higgs}$=100 and 1000 GeV.}
\label{fig:mt-mw}
\end{minipage}
}
\end{figure}

\noindent
Our measurement is in good agreement with Standard Model expectations, and is 
consistent with current measurements from W and Z production as well as from 
other neutrino experiments (Figures \ref{fig:mw}, \ref{fig:mt-mw}). The
data tend to collectively favor a light Higgs mass. The central value 
from recent global fits to all precision data is 
$M_{\rms Higgs}=98^{+57}_{-38}$ GeV with an upper bound of $M_{\rms Higgs} 
\leq 235$ GeV at 95\% CL \cite{higgs}. 

\section*{Beyond the Standard Model}
Outside the Standard Model, deviations between electroweak measurements in 
$\nu N$ scattering and those from other processes are sensitive to
new physics. In these proceedings we discuss two such possibilities:
neutrino oscillations and extra neutral vector gauge bosons.\\

%\subsection*{Neutrino Oscillations}
The presence of neutrino oscillations will directly shift our measured 
ratios, $R_{\rms meas}$, from their Standard Model predictions. Since 
$\nu_\mu$'s oscillating to either $\nu_e$'s or $\nu_\tau$'s would be less 
likely to produce a final state muon, we would expect to observe an excess 
of short events. Since no such excess is observed, single mode ($\nu$ or 
$\bar\nu$) limits can be set for both $\nu_\mu \rightarrow \nu_e$ 
and $\nu_\mu \rightarrow \nu_\tau$ oscillations. One advantage to this type 
of search is that the Paschos-Wolfenstein quantity, R$^{-}$, is particularly 
sensitive to CP-violating oscillations because it is formed from a difference 
in neutrino and antineutrino rates. As a result, NuTeV is presently the only 
experiment with direct limits on $\bar \nu_\mu \rightarrow \bar \nu_\tau$. 
More details on this analysis can be found elsewhere \cite{oscs}.\\

%\subsection*{Extra Neutral Vector Gauge Bosons}
Extra Z bosons (Z$^{'}$) are of interest not only because they are 
predicted by many Grand Unified Theories and superstring models, but also in 
light of recent experimental developments. Erler and Langacker have shown in
a recent global fit that the precision electroweak data are better described 
if an extra TeV-scale Z boson is included \cite{langacker}. Of course, a 
large portion of this improvement arises from the fact that the 2.5 $\sigma$ 
deviation of the new atomic parity violation (APV) measurement \cite{apv_exp} 
from the Standard Model prediction can be explained by including an additional
Z boson \cite{apv}.

\begin{figure}[ht]	
\mbox{
\begin{minipage}{0.5\textwidth}
\centerline{\epsfxsize 2.0 truein \epsfbox{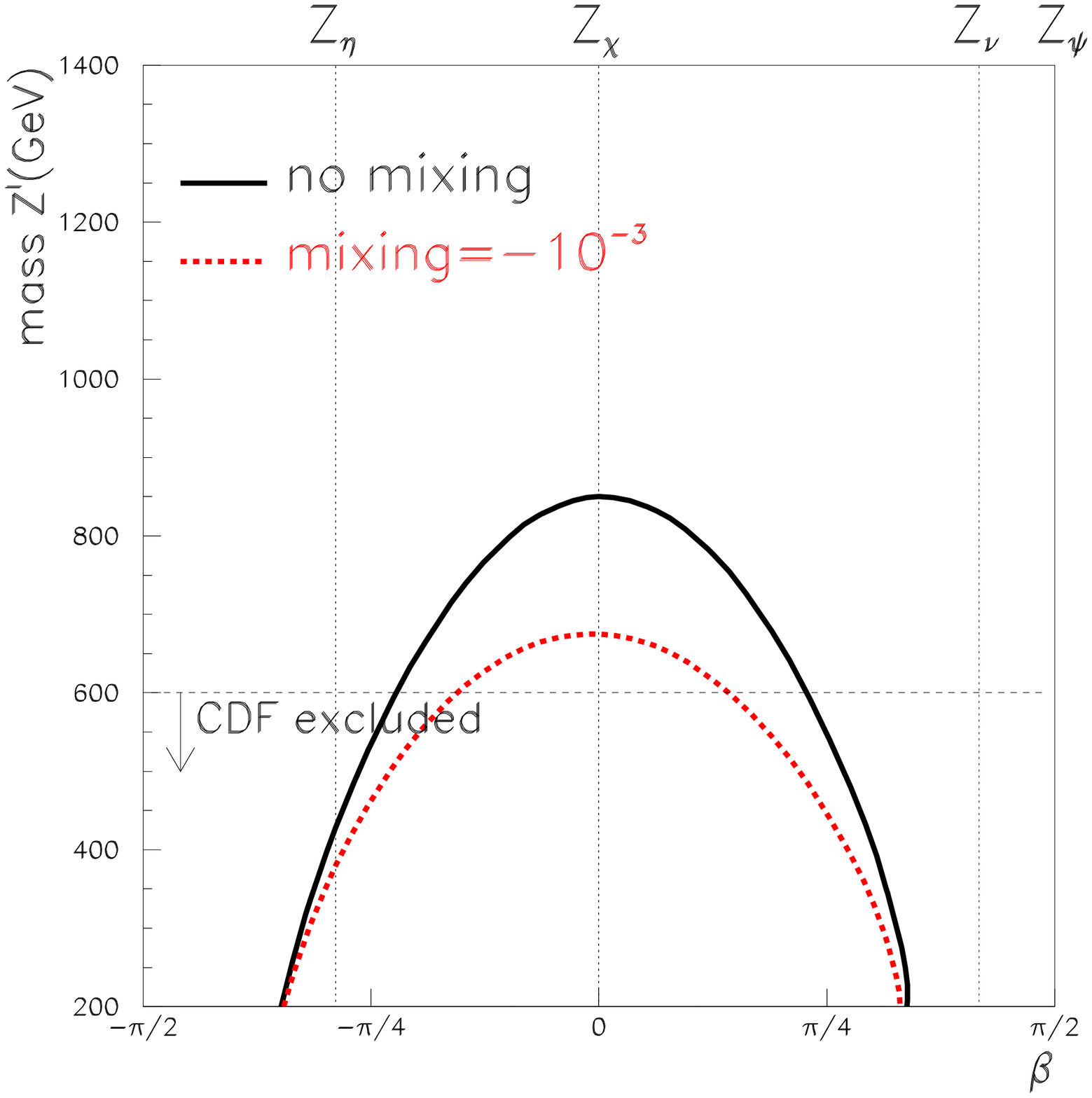}}   
\caption[]{NuTeV 95\% CL lower limits on the mass of the $Z^{'}$ (in GeV) as a 
           function of the $Z_{\chi}$, $Z_{\psi}$ mixing angle $\beta$. Limits 
           are shown for both no-mixing and allowed-mixing cases.} 
\label{fig:zprime1}
\end{minipage}\hspace*{0.02\textwidth}	
\begin{minipage}{0.5\textwidth}
\centerline{\epsfxsize 2.0 truein \epsfbox{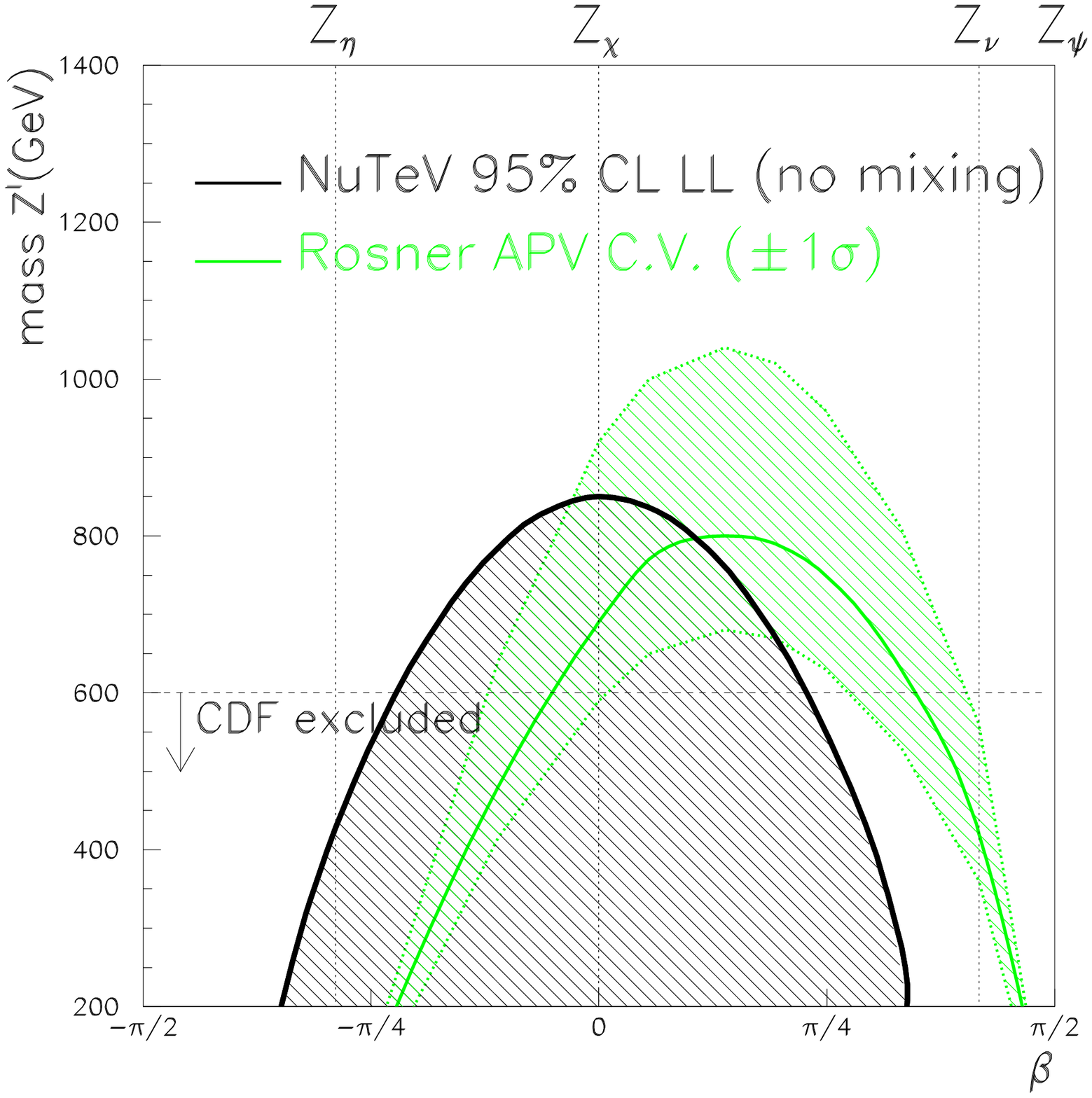}}   
\caption[]{Central value from Rosner's APV data analysis \cite{apv}. 
           NuTeV excludes the darker shaded region at 95\% CL. For purposes 
           of comparison, no Z-$Z^{'}$ mixing has been allowed.}
\label{fig:zprime2}
\end{minipage}
}
\end{figure}

In our case, extra Z bosons would manifest themselves as shifts in the 
neutrino-quark couplings away from their Standard Model values. These shifts 
can arise from both pure-Z$^{'}$ exchange as well as from Z-Z$^{'}$ mixing 
contributions. If we consider constraints on extra Z bosons in $E_6$ models, 
then these coupling shifts are well-determined \cite{shifts}. In this case, 
the lightest extra Z boson is a linear combination of the SO(10) singlet 
$Z_{\chi}$ and the SU(5) singlet $Z_{\psi}$:

\begin{equation}
           Z^{'} = Z_{\chi} \cos\beta + Z_{\psi} \sin\beta
\end{equation}

\noindent
expressed in terms of a free parameter $\beta$. Our data tend
to disfavor the inclusion of additional Z bosons, so we set a 95\% CL 
lower limit on the mass of such an extra $Z^{'}$ plotted as a function of 
$\beta$ (Figure \ref{fig:zprime1}). Limits are displayed for two cases: 
the $Z^{'}$ has no mixing with the Standard Model Z, and the 
more realistic case that allows for some level of mixing. For the latter, 
we input the Z pole data constraint that the level of Z-$Z^{'}$ mixing 
is small ($10^{-3}$) \cite{langacker}. Note that the inclusion of Z-$Z^{'}$ 
mixing, even at this small level, weakens the limits. At 95\% CL and assuming 
a 10$^{-3}$ level of Z-$Z^{'}$ mixing, we set lower mass limits of 675 and 
380 GeV for the $Z_{\chi}$ and $Z_{\eta}$ respectively. Direct searches have 
already excluded masses below $\sim$ 600 GeV \cite{cdf}. 

As can be seen from Figure \ref{fig:zprime1}, our maximum sensitivity is to 
the $Z_{\chi}$, and our exclusion peaks in the most viable region suggested 
by the APV data. Figure \ref{fig:zprime2} shows a direct comparison of the 
region NuTeV excludes in Z${'}$ mass-$\beta$ space to the favored central 
value from the APV data analyses \cite{apv}.

\section*{Conclusions}
NuTeV has successfully completed its data taking and has extracted a 
value of $\sin^2\theta_W$. The precision of this result represents 
a factor of two improvement over previous measurements in $\nu N$ 
scattering, because of reduced uncertainties associated with measuring the 
Paschos-Wolfenstein ratio, $R^{-}$. Interpreted within the framework of
the Standard Model, this result is equivalent to a determination of the 
W mass and is consistent with direct measurements of $M_{W}$. Outside
the Standard Model, this measurement can be used to set limits on neutrino 
oscillations as well as extra neutral vector gauge bosons.


\begin{references} 
\bibitem{ccfr} K.S. McFarland, {\em et al.}, Eur. Phys. Jour. {\bf C31}, 
               509 (1998).

\bibitem{pw} E.A. Paschos and L. Wolfenstein, Phys. Rev. {\bf D7}, 91 (1973).

\bibitem{detector} D.A. Harris, J. Yu, {\em et al.}, Nucl. Instr. Meth. 
                   {\bf A447}, 373 (2000).

\bibitem{dpfproc} K.S. McFarland, {\em et al.}, hep-ex/9806013; 
                  G. Zeller, {\em et al.}, hep-ex/9906037.

\bibitem{higgs} D.E. Groom, {\em et al.}, Eur. Phy. J. {\bf C15}, 1 (2000). 

\bibitem{oscs} D.A. Harris, {\em et al.}, submitted to the proceedings of 
               the International Europhysics Conference on High-Energy 
               Physics, Tampere Finland (1999).

\bibitem{langacker} J. Erler and P. Langacker, Phys. Rev. Lett. {\bf 84}, 212 
                    (2000).

\bibitem{apv_exp} S.C. Bennett and C.E. Wieman, Phys. Rev. Lett. {\bf 82},
                  2484 (1999).

\bibitem{apv} J. Rosner, hep-ph/9907524; 
              R. Casalbuoni, {\em et al.}, hep-ph/0001215.

\bibitem{shifts} P. Langacker, {\em et al.}, Rev. Mod. Phys. {\bf 64}, 87 
                 (1991)\\
                 G.-C. Cho, {\em et al.}, Nucl. Phys. {\bf B531}, 65 (1998).\\
                 D. Zeppenfeld and K. Cheung, hep-ph/9810277.
               
\bibitem{cdf} F. Abe, {\em et al.}, Phys. Rev. Lett. {\bf 79}, 2192 (1997).
   

\end{references}
\end{document}